# Mode structure and orbital angular momentum of spatiotemporal optical vortex (STOV) pulses

S.W. Hancock,  S. Zahedpour, and H.M. Milchberg

*Institute for Research in Electronics and Applied Physics*
*University of Maryland, College Park, MD 20742*

**Abstract**

We identify a class of modal solutions for spatio-temporal optical vortex (STOV) electromagnetic pulses propagating in dispersive media with orbital angular momentum (OAM) orthogonal to propagation. We find that symmetric STOVs in vacuum can carry half-integer intrinsic orbital angular momentum (OAM); for general asymmetric STOVs in a dispersive medium, the OAM is quantized in integer multiples of a parameter that depends on the STOV symmetry and the group velocity dispersion. Our results suggest that STOVs propagating in dispersive media are accompanied by a polariton-like quasiparticle. The modal theory is in excellent agreement with measurements of free space propagation of STOVs.

Several years ago, we reported the first measurement of spatiotemporal optical vortices (STOVs), which were found to emerge from the nonlinear self-focusing collapse arrest and filamentation of femtosecond optical pulses in air [1]. A STOV is a polychromatic electromagnetic structure with orbital angular momentum (OAM) and optical phase circulation defined in spacetime, with the OAM vector perpendicular to the direction of propagation. STOVs appear to underlie all self-focusing collapse scenarios, including relativistic self-focusing in plasmas and filamentation in transparent solids. More recently, we have demonstrated generation of STOV-carrying pulses using a $4f$ pulse shaper and studied their free space propagation from near field to far field, capturing their evolving spatio-temporal amplitude and phase [2], with this work later verified in experiments observing STOVs in the far field [3]. Our recent experiments demonstrating STOV OAM conservation in second harmonic generation verifies that STOV OAM applies at the single photon level [4-6].

The subject of optical vortices garnered renewed interest when Allen *et al.* [7] demonstrated a fundamental connection between intrinsic spatial OAM and Laguerre-Gaussian modes of integer topological charge $l$ : such beams carried OAM of $l\hbar$ per photon. In contrast to STOVs, the beams considered by Allen *et al.* are monochromatic, with the vortex axis, angular momentum, and linear momentum all aligned. Closer to our situation, the possibility of measuring spatiotemporal vortices has been suggested in [8] and dispersionless vortices have been considered in the spatiotemporal domain [9].

In this paper, we present a theoretical description of STOV-carrying pulses in both vacuum and in dispersive material media, with emphasis on their mode structure, propagation, and spatiotemporal orbital angular momentum. Recent theoretical work [10] has also considered spatiotemporal vortex pulses, but without a full modal analysis in dispersive media.

We start by looking for STOV-supporting modal solutions of the paraxial wave equation. To account for possible medium dispersion, we use the Fourier transformed wave equation for a uniform isotropic medium with dielectric function $\varepsilon(\omega)$ and wavenumber given by $k^2(\omega) = \omega^2 \varepsilon(\omega)/c^2$,

$$\left(\nabla_\perp^2 + \frac{\partial^2}{\partial z^2} + k^2(\omega)\right)\tilde{\mathcal{A}}(\mathbf{r}_\perp, z, \omega) = 0 , \tag{1}$$

where $\tilde{\mathcal{A}}$ is the $t \to \omega$ Fourier-transformed vector potential, pulse propagation is along $\hat{\mathbf{z}}$, $\mathbf{r}_\perp$ represents transverse coordinates orthogonal to $\hat{\mathbf{z}}$, and $\nabla_\perp^2$ is the corresponding transverse Laplacian. We assume $\tilde{\mathcal{A}}(\mathbf{r}_\perp, z, \omega) = \tilde{\mathbf{A}}(\mathbf{r}_\perp, z, \omega - \omega_0) e^{ik_0 z}$, where $\tilde{\mathbf{A}}$ is a slowly varying envelope and $k_0 = k(\omega_0)$ is the wavenumber at the central frequency. This yields $(\nabla_\perp^2 + 2ik_0\, \partial/\partial z)\, \tilde{\mathbf{A}} + (k^2(\omega) - k_0^2)\tilde{\mathbf{A}} = 0$ for $k_0|\partial\tilde{\mathbf{A}}/\partial z| \gg |\partial^2\tilde{\mathbf{A}}/\partial z^2|$. Using $k^2(\omega) - k_0^2 \approx 2k_0(k(\omega) - k_0)$ and expanding $k(\omega) = k_0 + k_0'(\omega - \omega_0) + k_0''(\omega - \omega_0)^2/2 + \cdots$ gives $2ik_0\, \partial\tilde{\mathbf{A}}/\partial z) = -\nabla_\perp^2 \tilde{\mathbf{A}} - 2k_0(k_0'\omega + k_0''\omega^2 + \cdots)\tilde{\mathbf{A}}$, where $k_0' = (\partial k/\partial\omega)_0 = v_g^{-1}$ is the inverse group velocity at $\omega_0$, and $k_0'' = (\partial^2 k/\partial\omega^2)_0 = \left(\partial v_g^{-1}/\partial\omega\right)_0$ is the group velocity dispersion (GVD). Assuming that the pulse bandwidth is not too large ($\Delta\omega/\omega_0 \ll 1$), keeping terms in the $k(\omega)$ expansion to second order is an excellent approximation. This gives, after transforming back to the time domain, $2ik_0\, \partial\mathbf{A}/\partial z = -(\nabla_\perp^2 + 2ik_0 k_0'\, \partial/\partial t - k_0 k_0''\, \partial^2/\partial t^2)\mathbf{A}$ where $\mathbf{A} = \mathbf{A}(\mathbf{r}_\perp, z, t)$. Finally, we make the substitutions $\xi = v_g t - z$ and $\beta_2 = v_g^2 k_0 k_0''$ to give

$$2ik_0 \frac{\partial}{\partial z} \mathbf{A}(\mathbf{r}_\perp, \xi; z) = (-\nabla_\perp^2 + \beta_2 \frac{\partial^2}{\partial \xi^2})\mathbf{A}(\mathbf{r}_\perp, \xi; z) = H\, \mathbf{A}(\mathbf{r}_\perp, \xi; z) . \tag{2}$$

Here, $\xi$ is a (local time-like) space coordinate in the frame of the pulse, $\beta_2$ is the dimensionless GVD, $H = (-\nabla_\perp^2 + \beta_2\, \partial^2/\partial \xi^2)$ is the spacetime propagator, and we separate $z$ with a semicolon as it plays the role of a time-like running parameter.

Next, we assume a uniformly polarized beam $\mathbf{A}(\mathbf{r}_\perp, \xi; z) = A(\mathbf{r}_\perp, \xi; z)\hat{\mathbf{e}}$, where $\hat{\mathbf{e}}$ is the complex polarization (here we take $\hat{\mathbf{e}} = \hat{\mathbf{y}}$ as in our experiments [2], where there are no effects of spin angular momentum [10]), and find modal solutions to Eq. (2) for $\mathbf{r}_\perp = (x, y)$:

$$A_{mpq}(x, y, \xi; z) = A_{mpq}^{(0)} u_m^x(x; z) u_p^y(y; z) u_q^\xi(\xi; z) , \tag{3}$$

where

$$u_q^\xi(\xi; z) = \frac{C_q}{\sqrt{w_\xi(z)}} H_q\left(\frac{\sqrt{2}\, \xi}{w_\xi(z)}\right) e^{-\xi^2/w_\xi^2(z)} e^{-ik_0 \xi^2/2\beta_2 R_\xi(z)} e^{i(q+1/2)\psi_\xi(z)} \tag{4a}$$

and

$$u_m^x(x; z) = \frac{C_m}{\sqrt{w_x(z)}} H_m\left(\frac{\sqrt{2}\, x}{w_x(z)}\right) e^{-x^2/w_x^2(z)} e^{ik_0 x^2/2R_x(z)} e^{-i(m+1/2)\psi_x(z)} . \tag{4b}$$

Here, $C_m = \left(\frac{2}{\pi}\right)^{\frac{1}{4}} (2^j\, m!)^{-\frac{1}{2}}$, $H_m$ is a Hermite polynomial of order $m$, $w_x(z) = w_{0x}(1 + (z/z_{0x})^2)^{1/2}$, $R_x(z) = z(1 + (z_{0x}/z)^2)$, $\psi_x(z) = \tan^{-1}(z/z_{0x})$, and $z_{0x} = k_0 w_{0x}^2/2$ is the $x$-based Rayleigh range. The expression for $u_p^y(y)$ is identical to Eq. 4b with the substitution $x \to y$ everywhere. Associated with $u_q^\xi(\xi; z)$ is $z_{0\xi} = k_0 w_{0\xi}^2/2|\beta_2|$, $w_\xi(z) = w_{0\xi}(1 + (z/z_{0\xi})^2)^{1/2}$, $R_\xi(z) = z(1 + (z_{0\xi}/z)^2)^{1/2}$, and $\psi_\xi(z) = \text{sgn}(\beta_2)\tan^{-1}(z/z_{0\xi})$. The quantities $w(z), R(z)$ and $\psi(z)$ express the $z$-variation in beam size, phase front curvature and Gouy phase shift as they do for standard transverse modes, except that here they also apply in the $\xi$ domain.

The "spot sizes" $w_{0x}$, $w_{0y}$, and $w_{0\xi}$ describe the transverse space and temporal shape of the beam envelope of the lowest order mode $((m,p,q) = (0,0,0))$ at $z = 0$, $A_{000}(x,y,\xi;z=0) = A_{000}^{(0)} e^{-(x^2/w_{0x}^2 + y^2/w_{0y}^2)} e^{-\xi^2/w_{0\xi}^2}$, which approximates the input beam to our pulse shaper. The effective wavenumber $k_0/\beta_2$ associated with $u_q^\xi(\xi)$ accounts for the different rate of spreading in temporal dispersion compared to transverse beam diffraction. We have allowed the beam to have elliptical envelopes in both the $x-y$ (space) and $x-\xi$ (spacetime) planes, and different phase curvatures in $x$, $y$, and $\xi$. The choice of HG basis functions for the solution of Eq. (2) is motivated by our experimental generation of STOV-carrying pulses using a $4f$ pulse shaper [2], which imposes rectilinearly-oriented ellipticity and astigmatism in both the space and spacetime domains.

We now consider propagation of the simplest STOV-carrying pulse generated by our pulse shaper, one with a spatiotemporal winding of topological charge $l = 1$ or $l = -1$. At $z = 0$, this pulse is constructed as

$$A_\alpha^{l=\pm 1}(x,y,\xi;z=0) = A_0 \left( \frac{\xi}{w_{0\xi}} \pm i \frac{x}{w_{0x}} \right) e^{-(x^2/w_{0x}^2 + y^2/w_{0y}^2)} e^{-\xi^2/w_{0\xi}^2}, \qquad (5)$$

As we will see, the spacetime eccentricity, $\alpha \equiv w_{0\xi}/w_{0x}$, is extremely important and will show up throughout these calculations. In the experiments, the $y$-direction is orthogonal to the pulse shaper grating rulings, and so after pulse reconstruction at the shaper output, the $y$-dependent envelope of the input pulse is reproduced [2].

In vacuum or in the very dilute medium (air) of the experiments of [2], $\beta_2 = 0$ and $v_g = c$, $u_q^\xi(\xi;z=0) = H_q(\sqrt{2}\xi/w_{0\xi})e^{-\xi^2/w_{0\xi}^2}$, and Eq.(5) can be represented as a linear combination of spacetime modes (Eqs. (4)) at $z = 0$:

$$A_\alpha^{l=\pm 1}(x,y,\xi;z=0) = A_0 u_0^y(y;0) \left( u_0^x(x;0) u_1^\xi(\xi;0) \pm i u_1^x(x;0) u_0^\xi(\xi;0) \right). \qquad (6)$$

Given this initial STOV field at $z = 0$, the propagator $H = (-\nabla_\perp^2 + \beta_2 \partial^2/\partial\xi^2)$ of Eq. (2) generates the full $z$-dependent evolution

$$A_\alpha^{l=\pm 1}(x,y,\xi;z) = A_0 u_0^y(y;z) \left( u_0^x(x;z) u_1^\xi(\xi;z) \pm i u_1^x(x;z) u_0^\xi(\xi;z) \right). \qquad (7)$$

For the case $w_{0x} = w_{0\xi}$ ($\alpha = 1$), the factor $u_0^x(x;z) u_1^\xi(\xi;z) \pm i u_1^x(x;z) u_0^\xi(\xi;z)$ is analogous to the superposition of the $0^{th}$ and $1^{st}$ order Hermite-Gaussian transverse modes ($HG_0$ and $HG_1$) to give the Laguerre-Gaussian spatial mode $LG_{space}^{0\pm 1} = HG_0(x)HG_1(y) \pm iHG_1(x)HG_0(y)$.

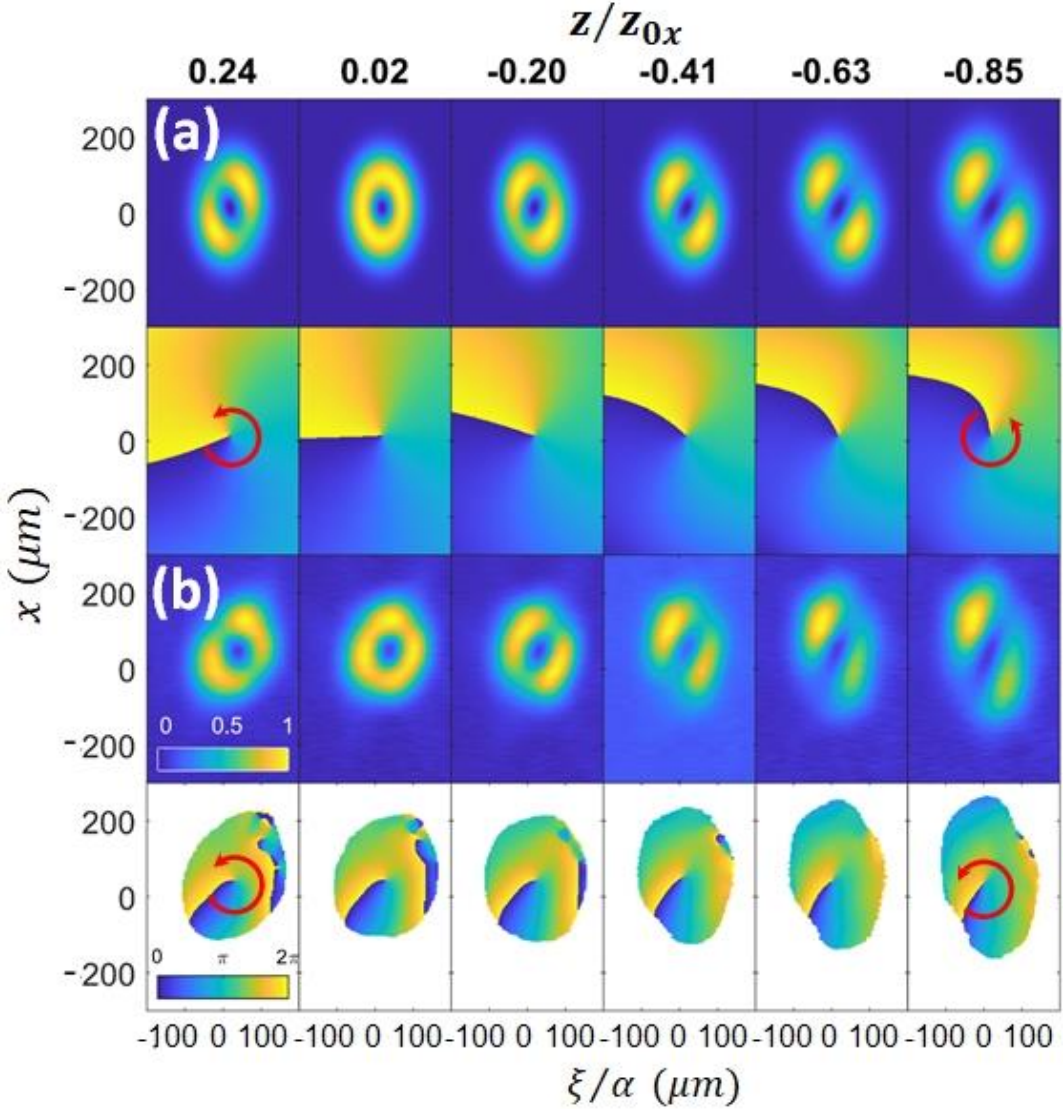

**Figure 1. (a)** Propagation evolution from $z/z_{0x} = -0.85$ to $0.24$ of the $l = 1$ STOV $A(x, y = 0, \xi; z)_{(l=+1)}$, plotted using the modal solution, Eq. (7). Top row: Normalized intensity $|A_\alpha^{l=+1}(x, y = 0, \xi; z)|^2$. Bottom row: Phase $\Phi(x, \xi) = \tan^{-1}[\text{Im}(a)/\text{Re}(a)]$, where $a = A_\alpha^{l=+1}(x, y = 0, \xi; z)$. The phase colourmap and red arrows show the direction of increasing phase $\Phi$. **(b)** Experiment: An $l = 1$ STOV is generated by passing a near-Gaussian pulse through a $4f$ pulse shaper with an $l = +1$ spiral phase plate at the shaper's Fourier plane (details in [2]), and the STOV amplitude and phase is captured in flight by TG-SSSI (details in [2, 11]). The experimental Rayleigh length is $z_{0x} = 46$mm. The horizontal axis for both (a) and (b) is normalized to the experimental spacetime eccentricity $\alpha = w_{0\xi}/w_{0x} = 0.3$. The phase plots in (b) are blanked out in regions of low intensity where phase extraction fails [7]. Within each panel, the pulse propagates from right to left.

Figure 1 compares theory and experiment, where Fig. 1(a) shows the amplitude and phase of $A_\alpha^{l=+1}(x, y = 0, \xi; z)$ from $z = -0.85z_{0x}$ to $z = 0.24z_{0x}$, computed with Eq. (7). It is seen that the field is a donut at the beam waist ($z = 0$) (as constructed) and evolves into spatio-temporally offset lobes with opposite spacetime tilt on either side of $z = 0$, with transverse diffractive spreading widening the beam. Here, we have used $\alpha = w_{0\xi}/w_{0x} = 0.3$ to match our measured eccentricity. The experimental results are shown in Fig. 1(b). To capture the in-flight amplitude and phase profiles of these pulsed spatiotemporal structures, we have employed a new diagnostic, transient-grating single-shot supercontinuum spectral interferometry (TG-SSSI), which is described in [11]. The measurements are in excellent agreement with our mode-based calculation, capturing the STOV field's evolution from a donut into spatiotemporally offset lobes, and matching the phase winding in each panel.

For a more direct analysis of angular momentum of STOVs, we now express our HG-based mode solutions in spacetime polar coordinates $(\rho, \Phi)$, where $x = \rho\sin\Phi$ and $\xi = \rho\cos\Phi$. Here, we can describe the spacetime phase winding by the topological charge $l$ and the single function $\Phi$, even for our general case of elliptical and astigmatic STOV pulses. The fundamental rectangular mode based on Eq. (3) is now written as

$$A_{000}(\rho, y, \Phi; z) = A_0 \sqrt{\frac{w_{0x}w_{0y}w_{0\xi}}{w_x(z)w_y(z)w_\xi(z)}} \exp\left(-\frac{\rho^2\sin^2\Phi}{w_x^2(z)} - \frac{y^2}{w_y^2(z)} - \frac{\rho^2\cos^2\Phi}{w_\xi^2(z)}\right)$$
$$\times \exp\left[ik_0\left(\frac{\rho^2\sin^2\Phi}{2R_x(z)} + \frac{y^2}{2R_y(z)} - \frac{\rho^2\cos^2\Phi}{2\beta_2 R_\xi(z)}\right)\right] \quad (8)$$
$$\times \exp\left[\frac{-i}{2}\left(\psi_x(z) + \psi_y(z) - \psi_\xi(z)\right)\right]$$

and the $l = \pm 1$ STOV pulse from our pulse shaper is

$$A_\alpha^{l=\pm 1}(\rho, y, \Phi; z) = A_{000}(\rho, y, \Phi; z)\left(\frac{\rho\cos\Phi}{w_\xi(z)}e^{i\psi_\xi(z)} \pm i\frac{\rho\sin\Phi}{w_x(z)}e^{-i\psi_x(z)}\right) \quad (9)$$

In our experiments, the $y$-dependent beam envelope shape, aside from transverse diffractive spreading, is preserved in propagation. So, we henceforth neglect $y$ variations in the beam by setting $y = 0$, noting that any 3D mode can be constructed by multiplying the $(x, \xi)$-dependent results by $u_n(y; z)$.

We now examine the STOV angular momentum, $\hat{\mathbf{y}}L_y$, which is orthogonal to the $x - \xi$ plane of spatiotemporal phase circulation. First, we must find the appropriate angular momentum operator $L_y$. To do so, we consider Eq. (2) along with the conservation of energy density flux $\mathbf{j}$ [12], $\partial |A|^2/\partial z = -\nabla \cdot \mathbf{j}$, where $\mathbf{j} = \mathbf{j}_\perp + \mathbf{j}_\parallel$, $\mathbf{j}_\perp = -i(2k_0)^{-1}(A^*\nabla_\perp A - A\nabla_\perp A^*)$ and $\mathbf{j}_\parallel = i\beta_2(2k_0)^{-1}[A^*(\partial/\partial \xi)A - A(\partial/\partial \xi)A^*]\hat{\boldsymbol{\xi}}$, where $\hat{\boldsymbol{\xi}}$ is unit vector along increasing $\xi$. This gives $\mathbf{j} = k_0^{-1}|A|^2(\nabla_\perp \Phi - \beta_2(\partial\Phi/\partial\xi)\hat{\boldsymbol{\xi}}) = k_0^{-1}|A|^2\nabla_{st}\Phi$, where $A = |A|e^{i\Phi}$ and $\nabla_{st} = \nabla_\perp - \hat{\boldsymbol{\xi}}\beta_2(\partial/\partial\xi)$ is the spacetime gradient. Therefore, the spacetime linear momentum operator is $\hat{\mathbf{p}} = -i\nabla_{st}$, giving $L_y = (-i\mathbf{r} \times \nabla_{st})_y = -i(\xi \partial/\partial x + x\beta_2 \partial/\partial \xi)$. In spacetime polar coordinates, this becomes

$$L_y = -i\left[\rho \sin\Phi \cos\Phi (1 + \beta_2)\frac{\partial}{\partial \rho} + (\cos^2\Phi - \beta_2 \sin^2\Phi)\frac{\partial}{\partial \Phi}\right] = L_y^e + L_y^i, \quad (10)$$

where we identify the first term as the extrinsic STOV angular momentum $L_y^e$, and the second term as the intrinsic STOV angular momentum $L_y^i$. Here, *intrinsic* refers to the origin-independent spatiotemporal angular momentum contribution, and *extrinsic* refers to the origin-dependent contribution which integrates to zero, $\langle L_y^e \rangle = 0$, when calculating the expectation value ($\langle \ \rangle$) of $L_y$ by integrating over $\rho$ ($0 \to \infty$) and $\Phi$ ($0 \to 2\pi$).

To calculate the STOV OAM associated with $A(\rho, y, \Phi; z)_{(l=\pm 1)}$, we note that it is sufficient to do so at the beam waist $z = 0$. This is because $\langle L_y \rangle$ is invariant with propagation, namely $(d/dz)\langle L_y \rangle = i(2k_0)^{-1}\langle [H, L_y] \rangle = 0$, owing to the fact that $[H, L_y] = 0$ ; $L_y$ commutes with the propagation operator. This procedure greatly simplifies the calculation, especially for non-zero $\beta_2$, where we consider the beam waist to be placed just inside the material interface ($z = 0^+$) without additional chirp from the material yet induced. At $z = 0$, Eq. (9) becomes

$$A_\alpha^{l=\pm 1} = A_\alpha^{l=\pm 1}(\rho, y = 0, \Phi; z = 0) = A_{000}(\rho, 0, \Phi; 0) \left( \frac{\rho \cos\Phi}{w_{0\xi}} \pm i \frac{\rho \sin\Phi}{w_{0x}} \right)$$
$$= A_0 \frac{\rho}{w_{0\xi}} \exp\left( -\frac{\rho^2}{2w_{0\xi}^2}(\cos^2\Phi + \alpha^2 \sin^2\Phi) \right) \left( \frac{(1 \pm \alpha)}{2} e^{i\Phi} + \frac{(1 \mp \alpha)}{2} e^{-i\Phi} \right) \quad (11)$$

This is nearly a linear combination of $LG^{0\pm 1}$ modes except for the $\Phi$-dependent exponential prefactor, which loses its angle dependence for $\alpha = 1$, yielding the symmetric spacetime Laguerre Gaussian mode $A_{\alpha=1}^{l=\pm 1} = LG_{spacetime}^{0\pm 1} = A_0 \, (\rho/w_{0\xi}) \exp(-\rho^2/2w_{0\xi}^2) \, e^{\pm i\Phi}$.

For arbitrary topological charge $l$, the $l^{th}$ order STOV pulse is

$$A_\alpha^l = A_\alpha^l(\rho, y = 0, \Phi; z = 0) \quad (12)$$
$$= A_0(\rho/w_{0\xi})^{|l|} \exp[-|l|\, (\rho^2/2w_{0\xi}^2)(\cos^2\Phi + \alpha^2 \sin^2\Phi)](\cos\Phi + i\alpha \, \text{sgn}(l) \sin\Phi \, )^{|l|} \quad .$$

For a STOV with a phase winding of charge $l$ and eccentricity $\alpha=1$, and for general $\alpha$,

$$\langle L_y \rangle_{l,\alpha=1} = \langle A_{\alpha=1}^l | L_y^i + L_y^e | A_{\alpha=1}^l \rangle = \langle A_{\alpha=1}^l | L_y^i | A_{\alpha=1}^l \rangle = \tfrac{1}{2}l(1 - \beta_2) \; , \quad (13a)$$

$$\langle L_y \rangle_{l,\alpha} = \langle A_\alpha^l | L_y^i + L_y^e | A_\alpha^l \rangle = \langle A_\alpha^l | L_y^i | A_\alpha^l \rangle = \tfrac{1}{2}l(\alpha - \beta_2/\alpha) \; , \quad (13b)$$

where $\langle L_y^e \rangle = 0$, and where $\langle L_y \rangle$ depends explicitly on topological charge $l$, STOV eccentricity $\alpha$ and material dispersion $\beta_2$.

This is a remarkable result, for which we will first consider the case $\alpha = 1$, a space-time symmetric STOV. For the case of vacuum ($\beta_2 = 0$), $\langle L_y \rangle = l/2$: STOV OAM *is quantized in half integer units*. For dispersive media ($\beta_2 \neq 0$), a quantum interpretation of the role of $\beta_2$ is strongly suggested, where one might consider the material disturbance induced by a STOV-encoded photon field as a new type of quasiparticle, a "STOV polariton".

A physical explanation for half-integer STOV orbital angular momentum in vacuum is that electromagnetic energy density flow in the pulse frame is purely along $\pm x$, or along $\nabla_\perp$. In our coordinates, for $l = +1$, energy density flows along $-x$ in advance of the STOV singularity and along $+x$ behind it, as seen in experiments and calculations in Fig. 1 and in ref. [2]. Because $\beta_2 = 0$ or is negligible in vacuum or extremely dilute media, there is no energy flow along $\xi$. This is in contrast to a standard $LG^{0\pm1}_{space}$ mode, where electromagnetic energy density circulates clockwise or counterclockwise around the singularity.

We now examine the physical meanings of $\beta_2$ and $\alpha$ in Eqs. (13). For simplicity, we first consider $\alpha = 1$ and later return to the interpretation of $\alpha$. Note that in Eq. (12) we used the vacuum STOV ($\beta_2 = 0$) –with its original spectral phase– implicitly at $z = 0^+$ (just inside the material) to calculate $\langle L_y \rangle_l$. But in reality, even at $z = 0^+$, the STOV spectral phase would have been modified by the dispersive material. Therefore, for a given phase winding $l$, the added term $-(l/2)\beta_2$ in $\langle L_y \rangle_l$, which is imposed by the full $\beta_2$-dependent $L^i_y$ operator (Eq. (10)), represents sharing of the pulse OAM with the material. This suggests that the material has an electromagnetic OAM response quantized in half integer steps of $\beta_2$ –we identify this object as a bulk medium "STOV polariton". For the case $\beta_2 = 1$, $\langle L_y \rangle_l = 0$ and the medium has apparently taken up $l/2$ units of angular momentum from the STOV field. It is interesting to note that $\beta_2 = 1$ for materials with a quadratic dispersion relation ($\omega \propto k^2$) or 'effective mass' for photons. This is a known dispersion dependence for polaritons [13, 14]. For a negatively dispersive material with $\beta_2 = -1$, we get $\langle L_y \rangle_l = l$, which we interpret as either that the STOV OAM is split between the photon and polariton field, or that the self-consistent electromagnetic object in the dispersive material has integer spatiotemporal OAM. Other values of $\beta_2$ give a range of OAM contributions between photons and polaritons.

We now address the physical meaning of $\alpha$. In the context of spatial OAM, there is nothing sacrosanct about circular symmetry except that the OAM for such beams coincides with the topological charge $l$ of the vortex [7]. References [13, 15] show that the intrinsic spatial OAM per photon of monochromatic beams with $w_{0x} \neq w_{0y}$ is determined by the ratio $w_{0y}/w_{0x}$. That is, the transverse beam shape is encoded onto the photon OAM. Although our spacetime paraxial wave equation (PWE) (Eq. (2)) is different than the spatial PWE, as is our STOV OAM operator, our conclusions regarding the STOV eccentricity parameter $\alpha = w_{0\xi}/w_{0x}$ are the same: $\alpha$ is encoded onto the intrinsic STOV OAM. In vacuum, STOV OAM is quantized in integer steps of $\alpha/2$ (or half-integer steps of $\alpha$), while in a dispersive medium, it is quantized in integer steps of $(\alpha - \beta_2/\alpha)/2$. For normally dispersive material with $\beta_2 > 0$, $\langle L_y \rangle_l = 0$ for $\alpha = \sqrt{\beta_2}$; the polariton most efficiently takes up STOV OAM when the eccentricity is tuned to the material's normalized GVD.

Considering the limit $\alpha \to 0$ in vacuum (and ignoring the breakdown in the slowly varying envelope approximation used to obtain Eq. (2)), $\langle L_y \rangle_l \to 0$ as is appropriate: the pulse loses the time-like contribution to its vorticity. In a dispersive medium, $\alpha \to 0$ corresponds to a shrinking temporal pulsewidth accompanied by increasing bandwidth, for which dispersion and the phase gradient contribution of $\hat{\xi}\beta_2 \partial\Phi/\partial\xi$ to $\langle L_y \rangle_l$ increase significantly. In classical terms, electromagnetic energy flow in $\hat{\xi}$ dominates that in $\hat{x}$. To be consistent with the *given* topological charge $l$, $|\langle L_y \rangle_l|$ must become large. For $\alpha \to \infty$, the pulse becomes very long and the effect of dispersion goes away ($\beta_2/\alpha \to 0$). Then, to be consistent for a given $l$, the phase gradient $\hat{x}\,\partial\Phi/\partial x$ must become very large, as does $\langle L_y \rangle_l$. In general, heuristic electromagnetic energy flow arguments like these provide good physical insight into the effects of varying $\alpha$ and $\beta_2$.

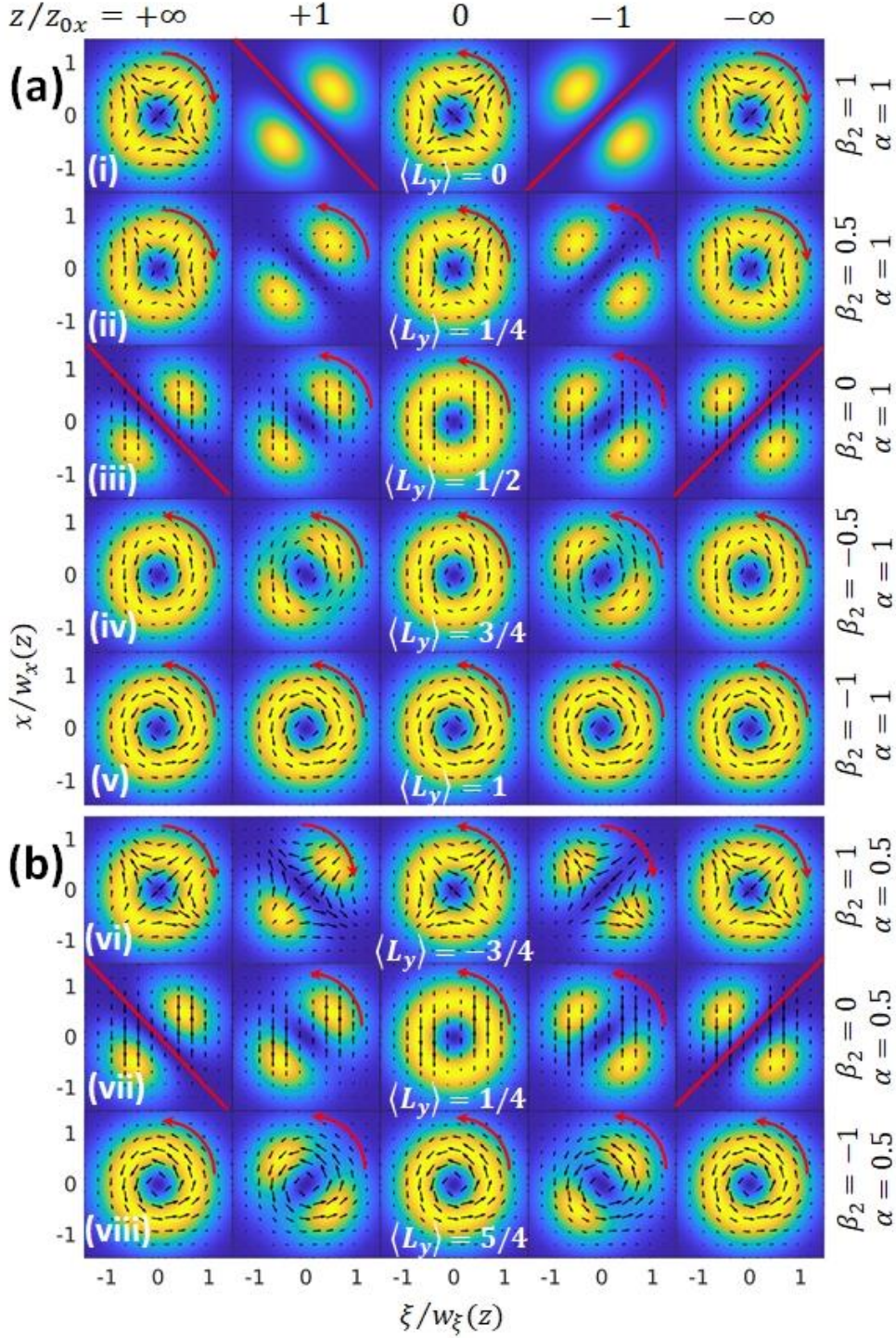

**Figure 2. (a)** Plots of STOV intensity $|A_{\alpha=1}^{l=1}(x, y = 0, \xi; z)|^2$ and energy density flux **j** (depicted by overlaid arrows), computed using Eq. (7) for $l = 1$, $\alpha = 1$ and $\beta_2 = 1, 0.5, 0, -0.5, -1$. The purely diffractive contributions to **j** have been subtracted out, leaving the flow contributing to OAM [17]. Propagation is shown through the beam waist ($z/z_{0x} = 0$) and into the far field ($\pm\infty$). The red arrows indicate the direction of spatiotemporal phase gradient $\nabla\Phi(x,\xi)$ and the red diagonals mark the boundary across which there is a phase jump of $\pi$. **(b)** Similar plots for spacetime-eccentric $l = 1$ STOVs with $\alpha = 0.5$ and $\beta_2 = 1, 0, -1$. For each row in (a) and (b), the value of $\langle L_y \rangle$ is shown in the $z = 0$ (centre) panels. Within each panel of (a) and (b), the pulse propagates right to left.

Figure 2(a) shows plots of STOV intensity $\left|A_{\alpha=1}^{l=1}(x, y = 0, \xi; z)\right|^2$ and energy density flux **j**, computed using Eq. (7), for $l = 1$, $\alpha = 1$ and $\beta_2 = 1,\ 0.5,\ 0,\ -0.5,\ -1$, and Fig. 2(b) shows similar results for a spacetime-eccentric $l = 1$ STOV with $\alpha = 0.5$. For each row of Fig. 2, $\langle L_y \rangle = \frac{1}{2}(\alpha - \beta_2/\alpha)$ is a constant. The purely diffractive contributions to **j** have been subtracted out, leaving the flow contributing to OAM [17]. The red arrows show the direction of the spatiotemporal phase gradient $\nabla \Phi(x, \xi)$, and the red diagonals mark the boundary across which there is a phase jump of $\pi$. In the panels with the red diagonal, even though the phase winding has disappeared, $\langle L_y \rangle$ remains at the constant value of that particular row. It is seen that for a STOV propagating in a medium with $\beta_2 > 0$, the energy density flow exhibits a 'saddle' pattern with respect to the singularity, while for $\beta_2 < 0$ the flow is spiral and, as discussed for $\beta_2 = 0$, the flow is restricted to $\pm x$. Note that for $\beta_2 = 1$, where $\langle L_y \rangle = \frac{1}{2}(1 - \beta_2) \propto \int dx d\xi (\mathbf{r} \times \mathbf{j})_y = 0$ and OAM is shared equally by the electromagnetic and polariton response, **j** vanishes everywhere at $z = z_{0x}$.

A range of interesting behaviour is observed in Fig. 2, with the main points summarized as follows: (1) In normally dispersive materials ($\beta_2 > 0$), the directions of the OAM and the phase gradient do not always coincide; the phase winding direction can flip to maintain OAM conservation (see rows (i),(ii), and (vi)); (2) The phase winding can disappear, yet nonzero $\langle L_y \rangle$ remains (rows (iii) and (vi)); (3) A donut-shaped STOV launched in vacuum or dilute media does not stay together as a donut; the spatio-temporal energy flow component $\mathbf{j}_\perp$ forces the donut into spatiotemporally offset lobes (rows (iii) and (vii)); (4) For $\beta_2 \neq 0$, the near and far field intensity profiles are self-similar (all rows except (iii) and (vii)); (5) There exists a self-similar STOV mode with integer OAM for $\alpha = 1$ and $\beta_2 = -1$ (row (v)). Classically, this is visualized as balanced STOV energy flow along $\hat{\mathbf{x}}$ and $\hat{\boldsymbol{\xi}}$.

In summary, we have presented an analysis for a new class of light states, spatio-temporal optical vortices (STOVs), with orbital angular momentum (OAM) orthogonal to propagation. In vacuum, these states are quantized in integer multiples of $\alpha/2$, where $\alpha$ is the STOV eccentricity parameter. For a symmetric STOV ($\alpha = 1$) in vacuum, the OAM is quantized in multiples of ½. In a dispersive medium, they are quantized in integer multiples of $(\alpha - \beta_2/\alpha)/2$, where $\beta_2$ is the normalized group velocity dispersion of the material, where we consider this OAM as shared between a photon and a STOV polariton. For $\alpha = 1$ and $\beta_2 = -1$, the STOV propagates as a self-similar mode with integer orbital angular momentum. We expect that our results will motivate further studies into the physics and applications of STOVs.

*Acknowledgements*. The authors thank Nihal Jhajj, who began this work, and Steve Rolston, Zakariah Chacko, and Dan Gordon for useful discussions. This work is supported by the Air Force Office of Scientific Research (FA9550-16-1-0121, FA9550-16-1-0284), the Office of Naval Research (N00014-17-1-2705, N00014-20-1-2233), and the National Science Foundation (PHY2010511).